\documentclass[%
 reprint,
 superscriptaddress,
 amsmath,amssymb,
 aps,
 prx,
 showkeys,
]{revtex4-2}

\usepackage[colorlinks = true,
            linkcolor = blue,
            urlcolor  = red,
            citecolor = red,
            anchorcolor = blue]{hyperref}
\usepackage{graphicx}
\usepackage{dcolumn}
\usepackage{bm}
\usepackage[mathlines]{lineno}
\usepackage{braket}
\usepackage{orcidlink}
\usepackage{ulem}

\usepackage{glossaries}

\newacronym{RIXS}{RIXS}{resonant inelastic x-ray scattering}
\newacronym{XAS}{XAS}{x-ray absorption spectroscopy}
\newacronym{ED}{ED}{exact diagonalization}
\newacronym{eph}{$e$-ph}{electron-phonon}
\newacronym{BKBO}{BKBO}{Ba$_{1-x}$K$_x$BiO$_3$}
\newacronym{BBO}{BBO}{BaBiO$_3$}
\newacronym{SNS}{SNS}{Spallation Neutron Source}
\newacronym{ORNL}{ORNL}{Oak Ridge National Laboratory}
\newacronym{NOMAD}{NOMAD}{Nano-scale Ordered Materials Diffractometer}
\newacronym{XRD}{XRD}{x-ray diffraction}
\newacronym{CDW}{CDW}{charge-density-wave}
\newacronym{PDF}{PDF}{pair distribution function}
\newacronym{XRF}{XRF}{x-ray fluorescence spectroscopy}
\newacronym{TFY}{TFY}{total fluorescence yield}
\newacronym{DFT}{DFT}{density functional theory}
\newacronym{PLD}{PLD}{pulsed laser deposition}
\newacronym{RHEED}{RHEED}{reflection high-energy electron diffraction}
\newacronym{DQMC}{DQMC}{determinant quantum Monte Carlo}
\newacronym{HMC}{HMC}{hybrid Monte Carlo}
\newacronym{PES}{PES}{photoemission spectroscopy}
\newacronym{PDOS}{PDOS}{partial density of states}
\newacronym{DOS}{DOS}{density of states}
\newacronym{SSH}{SSH}{Su-Schrieffer-Heeger}
\newacronym{FWHM}{FWHM}{full width at half maximum}
\newacronym{ARPES}{ARPES}{angle-resolved photoemission spectroscopy}

\usepackage[
]{todonotes}

\begin{document}

\preprint{}

\title{Persistence of small polarons into the superconducting doping range of \texorpdfstring{{Ba$_{1-x}$K$_x$BiO$_3$}}{Ba1-xKxBiO3}}

\author{Muntaser Naamneh\orcidlink{0000-0001-6676-9179}} 
\affiliation{Department of Physics, Ben-Gurion University of the Negev, Beer-Sheva, 84105, Israel}
\author{Eric~C.~O'Quinn}
\affiliation{Department of Nuclear Engineering, The University of
Tennessee, Knoxville, TN 37996, USA}
\author{Eugenio Paris}
\affiliation{Center for Photon Science, Paul Scherrer Institut, CH-5232 Villigen PSI, Switzerland}
\author{Daniel McNally}
\affiliation{Center for Photon Science, Paul Scherrer Institut, CH-5232 Villigen PSI, Switzerland}
\author{Yi Tseng}
\affiliation{Center for Photon Science, Paul Scherrer Institut, CH-5232 Villigen PSI, Switzerland}
\author{Wojciech~R.~Pude\l{}ko}
\affiliation{Center for Photon Science, Paul Scherrer Institut, CH-5232 Villigen PSI, Switzerland}
\author{Dariusz~J.~Gawryluk}
\affiliation{Laboratory for Multiscale Materials Experiments, Paul Scherrer Institut, CH-5232 Villigen PSI, Switzerland}
\author{Jacob Shamblin}
\affiliation{Department of Nuclear Engineering, The University of
Tennessee, Knoxville, TN 37996, USA}
\author{Benjamin~Cohen-Stead\orcidlink{0000-0002-7915-6280}}
\affiliation{Department of Physics and Astronomy, The University of Tennessee, Knoxville, TN 37996, USA}
\affiliation{Institute of Advanced Materials and Manufacturing, The University of Tennessee, Knoxville, TN 37996, USA\looseness=-1} 
\author{Ming~Shi}
\affiliation{Center for Photon Science, Paul Scherrer Institut, CH-5232 Villigen PSI, Switzerland}
\affiliation{Center for Correlated Matter and School of Physics, Zhejiang University, 310058, Hangzhou, China}
\author{Milan Radovic}
\affiliation{Center for Photon Science, Paul Scherrer Institut, CH-5232 Villigen PSI, Switzerland}
\author{Maik K.~Lang}
\affiliation{Department of Nuclear Engineering, The University of
Tennessee, Knoxville, TN 37996, USA}
\author{Thorsten Schmitt}
\affiliation{Center for Photon Science, Paul Scherrer Institut, CH-5232 Villigen PSI, Switzerland}
\author{Steven Johnston\orcidlink{0000-0002-2343-0113}}
\email{sjohn145@utk.edu}
\affiliation{Department of Physics and Astronomy, The University of Tennessee, Knoxville, TN 37996, USA}
\affiliation{Institute of Advanced Materials and Manufacturing, The University of Tennessee, Knoxville, TN 37996, USA\looseness=-1} 
\author{Nicholas C. Plumb\orcidlink{0000-0002-2334-8494}}
\email{nicholas.plumb@psi.ch}
\affiliation{Center for Photon Science, Paul Scherrer Institut, CH-5232 Villigen PSI, Switzerland}

\date{\today}

\begin{abstract}
    Bipolaronic superconductivity is an exotic pairing mechanism proposed for materials like \gls*{BKBO}; however, conclusive experimental evidence for a (bi)polaron metallic state in this material remains elusive. Here, we combine resonant inelastic x-ray and neutron total scattering techniques with advanced modelling to study the local lattice distortions, electronic structure, and \gls*{eph} coupling in \gls*{BKBO} as a function of doping. Data for the parent compound ($x = 0$) indicates that the electronic gap opens in predominantly oxygen-derived states strongly coupled to a long-range ordered breathing distortion of the oxygen sublattice. Upon doping, short-range breathing distortions and sizable \gls*{eph} coupling persist into the superconducting regime ($x = 0.4$). Comparisons with exact diagonalization and determinant quantum Monte Carlo calculations further support this conclusion. Our results provide compelling evidence that \gls*{BKBO}'s metallic phase hosts a liquid of small (bi)polarons derived from local breathing distortions of the lattice, with implications for understanding the low-temperature superconducting instability.
\end{abstract}


\maketitle

\glsresetall

Unconventional superconductivity typically appears near a parent phase with an ordered insulating or poorly metallic ground state. Notable examples include antiferromagnetic Mott insulating order in the cuprates~\cite{Keimer2015from}, spin-density-wave order in the Fe-based superconductors~\cite{Stewart2011superconducticity}, charge order in kagome superconductors~\cite{Plokhikh2024}, and \gls*{CDW} order in the bismuthates~\cite{Sleight2015} and antimonates~\cite{Kim2022superconductivity}. Understanding the parent compounds and their residual correlations upon doping are key steps toward identifying the pairing mechanisms in these materials. Addressing this question is challenging, however, as the parent compounds' correlations by themselves are complex, and various degrees of freedom can become intertwined as the materials are doped, leading to the emergence of competing or coexisting states~\cite{Fradkin2015} and novel quasiparticles.

Perovskite bismuth oxides~\cite{Cava1988superconductivity, Sleight2015} are a promising platform to disentangle the roles of various interactions in oxide superconductors. The superconductivity in these compounds can be regarded as ``unconventional'' in the sense that it emerges out of an insulating parent phase, with a doping-dependent transition temperature, $T_\mathrm{c}$, that can reach over 30 K. Moreover, their normal state resistivity follows a ``strange metal''-like linear temperature dependence that becomes more Fermi liquid-like at high doping \cite{Nagata1999} --- a common feature of several correlated electron systems, including the cuprates~\cite{Ayres2021incoherent}. 
The temperature dependence of the specific heat at $T_c$ also departs from expectations of the Bardeen-Cooper-Schrieffer theory \cite{Kuentzler1991superconductivity}. At the same time, superconductivity in the bismuth oxides appears to be phonon-mediated with an $s$-wave order parameter \cite{Kondoh1989, Zhao2000, Snezhko2004}. 
The compounds do not exhibit any magnetic order \cite{Tian}, and measured \gls*{BKBO} band structures closely match \gls*{DFT} calculations, implying that short-range Hubbard-like electronic correlations can be neglected in certain contexts~\cite{Plumb2016, Wen_ARPES_BKBO}. This view is further supported by recent \gls*{HMC} calculations for the parent compound  \gls*{BBO}, which reproduce quantitative details of its \gls*{CDW} state using an uncorrelated DFT-derived tight-binding model and coupling to the Bi-O bond-stretching modes~\cite{CohenStead2023}. Thus, in the bismuthates, much of the phenomenology associated with unconventional superconductors appears to arise from a relatively limited set of interactions, and \gls*{BKBO} presents a unique opportunity to investigate aspects related to \gls*{eph} coupling, polaron formation, and superconductivity. 

\begin{figure*}[t]
    \includegraphics[width=\textwidth]{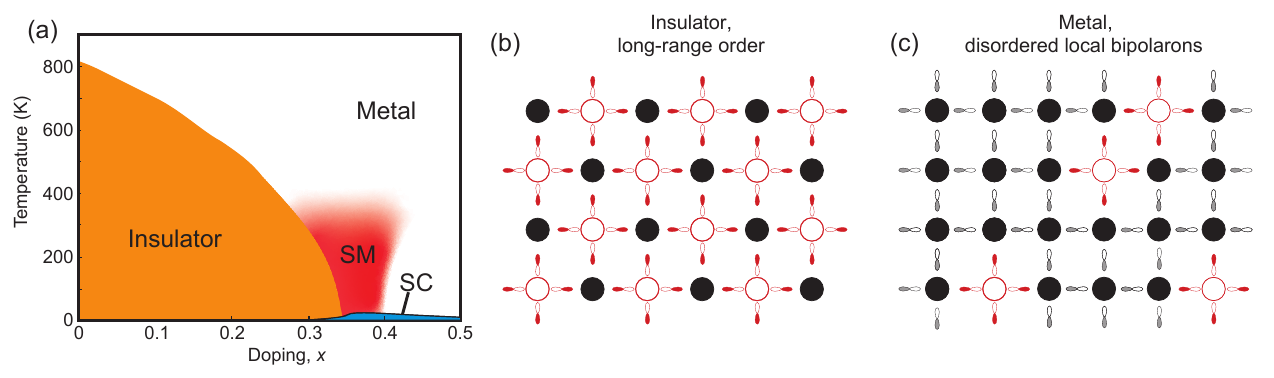} 
	\caption{Overview of Ba$_{1-x}$K$_x$BiO$_3$. 
    (a) Sketch of the doping-temperature ($x$ - $T$) phase diagram. The superconducting phase is labeled SC, and the strange metal-like behavior above the superconducting state is labeled SM. (b) Two-dimensional sketch of the long-range ordered oxygen breathing distortions in the insulating parent compound ($x=0$). Sites where the O $2p$ orbitals collapse around Bi $6s$, forming bipolarons, are shown in red. 
    (c) Scenario of short-range, disordered (bi)polaron breathing distortions persisting in the metallic phase.}
	\label{fig:fig1}
\end{figure*}

The phase diagram of \gls*{BKBO} is shown in Fig.~\ref{fig:fig1}(a). The parent compound is a semiconductor with an indirect band gap of 0.25~eV~\cite{Hellman1990}. It has a perovskite structure with a long-range ``breathing'' structural distortion in which the BiO$_6$ octahedra are alternately expanded and collapsed following a $\bm{Q} = (\pi,\pi,\pi)/a$ wave vector ($a$ is the Bi-Bi bond distance), as sketched in Fig.~\ref{fig:fig1}(b). At $x=0$, the system is half-filled, with charges coupled to the breathing distortion to form a checkerboard of trapped holes or (bi)polarons~\cite{Franchini2009, CohenStead2023}. The polarons are diluted upon hole doping, and superconductivity emerges at a doping level of $x \approx 0.3$. The maximum or optimal $T_\mathrm{c}$ occurs near or slightly below $x=0.4$.

Neither the details of the insulator-superconductor transition nor the polarons' evolution with doping, along with their impact on the superconducting state, are clear. The current understanding suggests two possible scenarios to explain how superconductivity emerges in the bismuthates. The first scenario minimizes the significant influence of the breathing distortion of the parent compound state on the superconducting state. Instead, it embraces a conventional mechanism where pairing is mediated by \gls*{eph} coupling that is enhanced by long-range Coulomb interactions~\cite{Yin2013correlation, Wen_ARPES_BKBO}. An alternative perspective posits that the small polarons derived from local breathing distortions of the BiO$_6$ octahedra persist in the metallic state, as sketched in Fig.~\ref{fig:fig1}(c), creating a (bi)polaronic liquid that ultimately forms a superconducting condensate at low temperatures~\cite{Khazraie2018bond, Khazraie2018oxygen, Li2020quantum, Jiang2021polaron}. While some studies have found evidence supporting the existence of such a polaron liquid \cite{Pashkevich2000, Menushenkov2003, Sugai1985, Tajima1987, Heald1989, Menushenkov2000, Kim2003, Braden2002}, others have cast doubt on this scenario \cite{McCarty1989, Salem-Sugui1991, Tajima1992, Greven2023}. 

Here, we investigate the presence and evolution of small polarons in \gls*{BKBO} across its insulator-superconductor transition using two experimental techniques: neutron scattering with \gls*{PDF} analysis to probe the crystal structure and O $K$-edge \gls*{RIXS} to probe the \gls*{eph} coupling across the transition. The experiments are complemented by determinant quantum Monte Carlo (DQMC) and exact diagonalization (ED) calculations. In contrast to many other spectroscopic or diffraction techniques, the \gls*{PDF} and \gls*{RIXS} methods used here are sensitive to the local environments of the BiO$_6$ octahedra and the oxygen sublattice. The neutron \gls*{PDF} measures the distribution of bond lengths in the system; it is thus sensitive to the presence of local lattice distortions, even if they are highly disordered and hence ``invisible'' to standard diffraction techniques. Meanwhile, \gls*{RIXS}, by virtue of being performed on the oxygen $K$-edge resonance, can be tuned to select specifically those sites where the oxygen states are most strongly coupled to lattice breathing mode distortions and provide information about the coupling strength~\cite{Ament2011, Lee2013, Devereaux2015, Bieniasz2021, thomas2024theory}. Together, the experiments and simulations show a consistent picture in which breathing distortions persist across the insulator-superconducting transition and remain strongly coupled to the electrons. Our findings contrast with a recent claim that these distortions vanish at low doping before the emergence of superconductivity \cite{Greven2023}.

\section*{Results}

\begin{figure*}[t]
    \centering
    \includegraphics[width=0.90\textwidth]{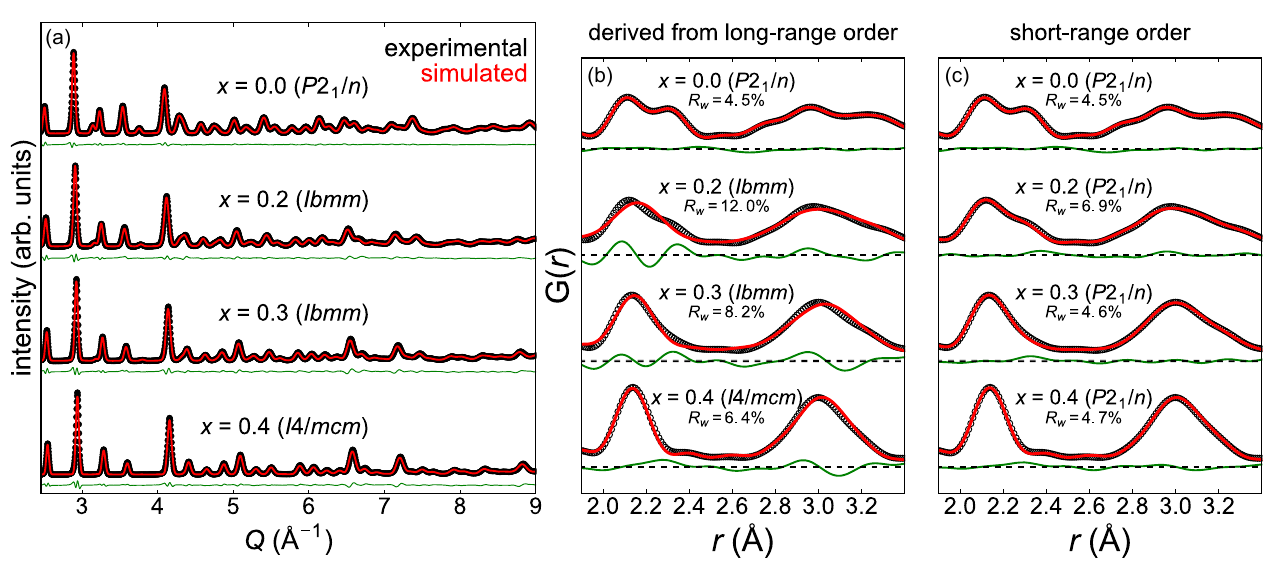}
    \caption{Rietveld and small-box refinements of neutron diffraction data.  
    (a) Rietveld refinement of neutron diffraction patterns. Data was taken at $T = 100~\mathrm{K}$ for Ba$_{1-x}$K$_x$BiO$_3$ samples with different doping levels. Black dots correspond to the measurements. Simulated diffraction patterns for each structure are shown in red. Green curves are the differences between the data and simulations.
    (b), (c) Small-box refinements of neutron total scattering data. Results are collected from Ba$_{1-x}$K$_x$BiO$_3$ at 100 K and fit against (b) the structural model inferred from Rietveld refinement and (c) the monoclinic $P2_1/n$ model, which contains two distinct Bi sites. All refinements were performed over the range 
    $1.5 \le r \le 10$ \AA.
    }
    \label{fig:refinement}
\end{figure*}

\subsection*{Local structure characterization}
The neutron scattering measurements were performed on polycrystalline \gls*{BKBO} samples, as described in the Methods section.  We begin by analysing \gls*{BKBO}'s average crystal structure when viewed over long length scales, which is provided by a traditional diffraction analysis based on Rietveld refinement. In this case, \gls*{BKBO} is expected to have different (average) crystal structures and symmetries depending on doping level $x$ and temperature $T$~\cite{Sleight2015}. The parent compound adopts monoclinic structures in space group $P2_1/n$ or space group $I2/m$ at low ($T < 140~\mathrm{K}$) and high temperatures ($140~\mathrm{K} < T < 430~\mathrm{K}$), respectively, while orthorhombic structures in space group $Ibmm$ appear for intermediate doping levels ($x=0.2,0.3$), and tetragonal structures in either space group $I4mm$ or $I4/mcm$ form in the near-optimal doping range for superconductivity ($x = 0.4$). To confirm this, we performed Rietveld refinements on all neutron diffraction patterns, as shown in Fig.~\ref{fig:refinement}(a). As expected, the diffraction pattern of the parent compound is modeled best using the monoclinic structure ($P2_1/n$). The data in the intermediate doping region ($x = 0.2$ and $0.3$) are fit well with the orthorhombic structure ($Ibmm$), while the data for the $x=0.4$ sample is fit well with the higher-symmetry tetragonal structure ($I4/mcm$). 

To investigate the evolution of the local crystal structure with doping, we converted the neutron scattering data into \glspl*{PDF}, $G(r)$, following the Fourier transform in Eq.~\eqref{eq:fourier}. The results at $T = 100~\mathrm{K}$ are shown in Figs.~\ref{fig:refinement}(b) and \ref{fig:refinement}(c), where we focus on distances $1.9\le r \le 3.3~\text{\AA}$. (Supplementary Note 1 presents the same data over wider distance ranges \cite{SuppMat}.) 
Two features are clear in the PDF of all compositions evaluated: peaks corresponding to Bi-O pair correlations ($2.0\le r \le 2.4~\text{\AA}$) and peaks corresponding to a convolution of Ba/K-O and O-O pair correlations ($2.7\le r \le 3.3~\text{\AA}$). Notably, the Bi-O correlation is bimodal for the $x = 0.0,$ $0.2,$ and $0.3$ samples, while appearing to converge to a single peak for the $x = 0.4$ sample despite the $I4/mcm$ average structure of this composition permitting two distinct Bi-O distances \textit{per} octahedron, with all octahedra equivalent.
While the orthorhombic and tetragonal structures inferred from the diffraction analysis reasonably describe the local structure for the intermediate-doped ($x = 0.2$, 0.3) and higher-doped samples that exhibit superconductivity ($x = 0.4$), respectively (see Fig.~\ref{fig:refinement}(b)), the \gls*{PDF} can be better modeled using the monoclinic phase ($P2_1/n$) across the whole doping range, as quantified by the fit goodness $R_w$ (see Fig.~\ref{fig:refinement}(c)).

\begin{figure}[h]
    \centering
    \includegraphics[width=\columnwidth]{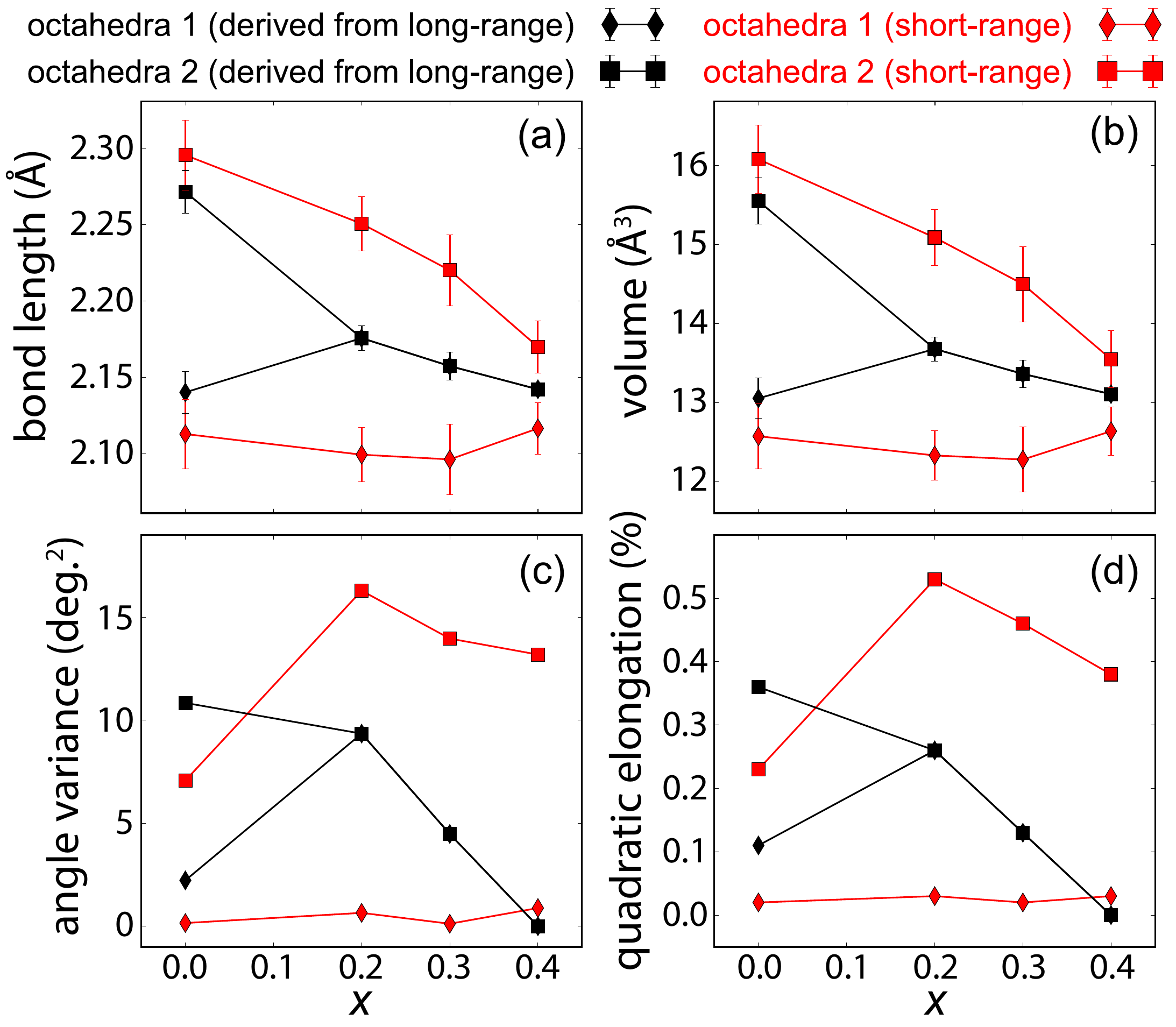}
    \caption{Evolution of the local BiO$_6$ octahedra with doping.
    Each panel shows the evolution of different local structural parameters as inferred from Rietveld refinement (black) of the diffraction patterns and small-box refinement (red) of PDFs collected at 100 K. Results are shown for (a) the local Bi-O bond lengths, (b) BiO$_6$ octahedral volumes, (c) Bi-O bond angle variance, and (d) quadratic elongation of the BiO$_6$ octahedra. 
    }
    \label{fig:polyhedra_figure}
\end{figure}

To better understand the nature of the short-range distortions, Fig.~\ref{fig:polyhedra_figure} investigates the refined local structure obtained from the different structural models in more detail. A key feature of the monoclinic local structure is that it allows for two inequivalent octahedral sites occupied by Bi (Wyckoff positions $2a$ and $2b$) and thus for any potential breathing distortion. In contrast, the long-range tetragonal and orthorhombic structure only allows for one octahedral site. 
Using these models, we analyzed the local Bi-O lengths (Fig.~\ref{fig:polyhedra_figure}(a)) and octahedral volumes (Fig.~\ref{fig:polyhedra_figure}(b)), which are indicative of the breathing distortion. We also examined the octahedral bond angle variance (Fig.~\ref{fig:polyhedra_figure}(c)), which indicates trigonal distortion, and quadratic elongation (Fig.~\ref{fig:polyhedra_figure}(d)), a dimensionless quantity independent of the effective size of the octahedra, that is correlated with tetragonal distortion~\cite{Robinson1971quadratic}.

Small-box refinements of the \glspl*{PDF} based on the {\textit{P}2\textsubscript{1}/\textit{n}} monoclinic structure, which includes two inequivalent octahedra, identify two distinct sets of Bi-O bond lengths and BiO$_6$ volumes for each composition evaluated. This model provides a better description of the local structure than other models with a greater number of Bi-O distances \textit{per} octahedron but only a single octahedron \textit{per} unit cell (see Supplementary Note 1 \cite{SuppMat}). We find a smaller octahedron (red diamonds) exhibiting very little change in volume across the series, while a larger octahedron (red squares) decreases in volume monotonically with $x$. The two octahedra also display distinct characteristics in terms of their bond angle variances and quadratic elongations. For all measured dopings, both of these quantities are consistently near zero for the smaller octahedron and clearly finite for the larger octahedron. The smaller BiO$_6$ unit therefore maintains near-ideal octahedral shape while the larger one possesses both trigonal and tetragonal distortions.

The local structure obtained from the \gls*{PDF} analysis contrasts with the structure averaged over long distances, as determined by Rietveld refinement. In the Rietveld analysis (black squares), the $x=0.0$ sample exhibits two distinct octahedra that not only have different volumes but also their own unique, finite angle variances and quadratic elongations. At $x \ge 0.2$, these merge into a single octahedron whose volume, angle variance, and quadratic elongation all decrease as $x$ further increases. The trigonal and tetragonal distortions signified by the angle variance and quadratic elongation eventually vanish at $x = 0.4$. 

The two diffraction analyses combine to tell a nuanced but self-consistent story: averaged over long length scales, the two inequivalent, trigonally, and tetragonally distorted Bi octahedra ($x=0$) merge into one for $x>0.2$ and trend toward ideal geometry at $x=0.4$. However, across short length scales, two inequivalent Bi octahedra persist in all compositions: a smaller, nearly idealized octahedron and a larger one that consistently exhibits trigonal and tetragonal distortions. Therefore, the distinction between the two local Bi octahedra is not simply in the volume variance between the two but also the difference of distortions. 
These trends are also observed in the PDF data recorded at $T = 300$ K data (see Supplementary Note 1 \cite{SuppMat}), indicating that the local distortions inferred here persist up to at least room temperature. Supplementary Note 1 also provides further discussion on the robustness and uniqueness of the short-range model derived here.

\begin{figure}[t]
    \centering
    \includegraphics[width=\columnwidth]{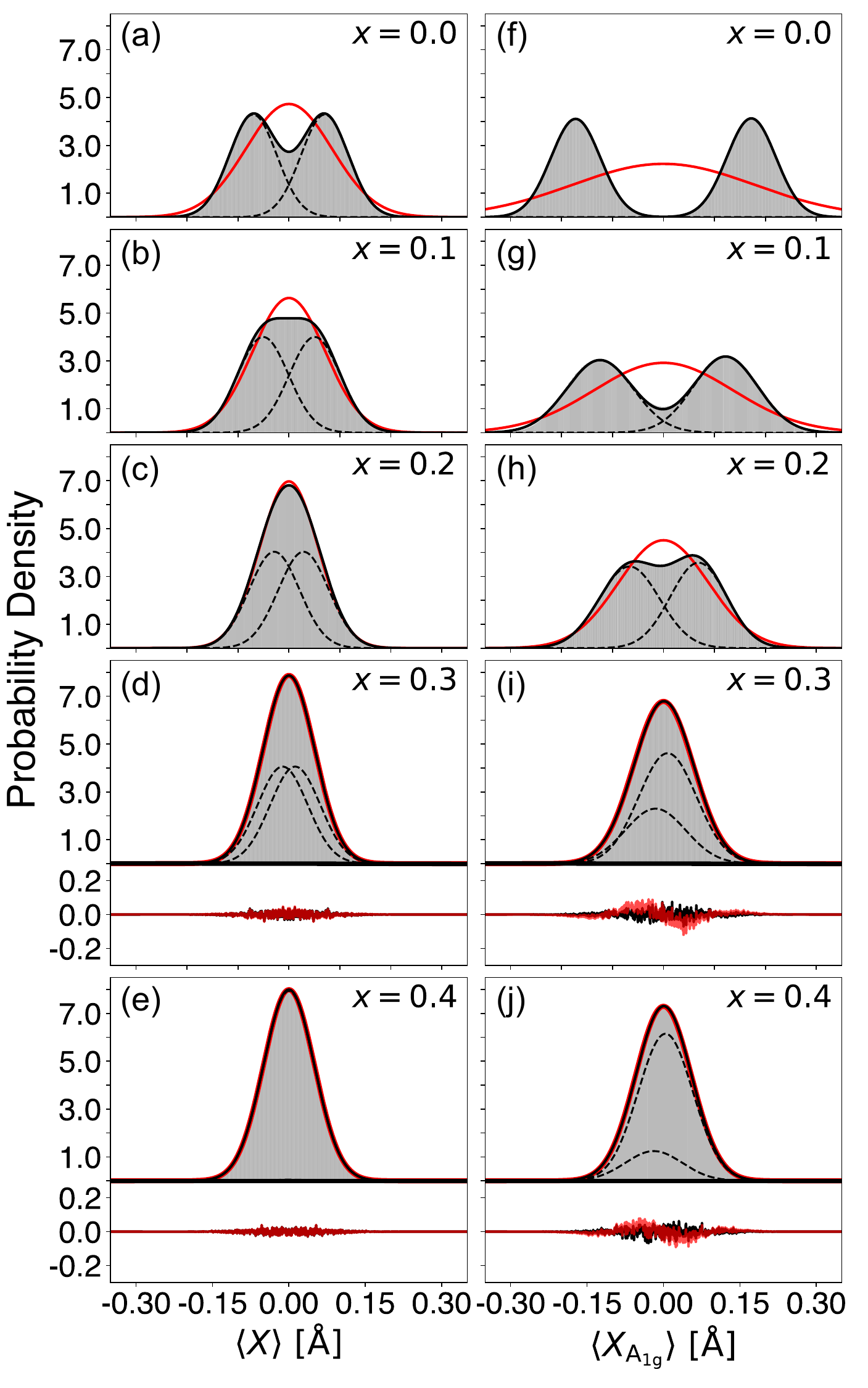}
    \caption{DQMC simulations of the Bi-O bond-lengths as a function of doping.
    (a)--(e) Probability distributions of the lattice displacements measured in our DQMC simulations of a cubic Bi$_{64}$O$_{256}$ cluster at $T = 300$~K. Data for different excess hole concentrations $x = \langle n\rangle-1$ (in holes/Bi) is shown, as indicated in each panel. 
    (f)--(j) The same data but projected on the $A_\mathrm{1g}$ breathing mode. The gray shaded area in each panel is the measured distribution. The dashed black and solid red lines are one- and two-Gaussian fits to the data, respectively. The black dashed curves show the individual contributions of the peaks within the two-Gaussian fits. The bottom portions of panels (d), (e), (i), and (j) plot the residuals between the fits and the data to better highlight the differences between the one- and two-Gaussian models (red and black, respectively). 
    }
    \label{fig:DQMC}
\end{figure}

\begin{figure*}[t]
	\includegraphics[width=1\textwidth]{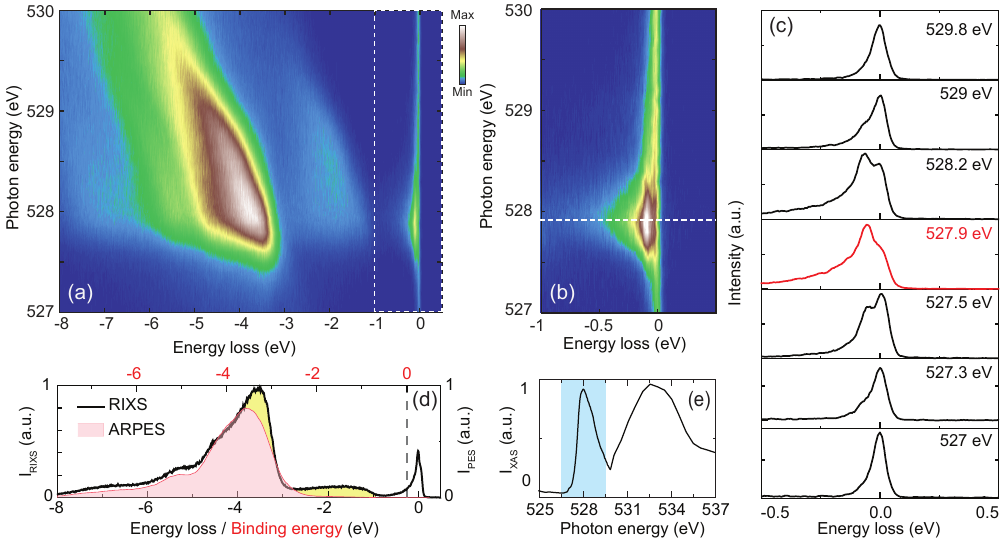} 
	\caption{RIXS data from a BBO thin film.
    (a) RIXS intensity map of BBO across the O $K$-edge as highlighted by blue in panel (e). 
    (b) A close-up view of the dashed rectangular region in the RIXS map of (a), highlighting the low-energy excitations. The dashed line marks the resonance energy.
    (c) RIXS spectra near the elastic peak, plotted for various incident energies. The series displays an intensity enhancement of the low-energy excitations at the resonance energy of 527.9 eV (red curve), which corresponds to the peak of the O $K$-edge absorption curve illustrated in panel (e). 
    (d) Fluorescence measured by RIXS at the resonance energy (solid black line, left and bottom axes) plotted along with an angle-integrated photoemission spectrum (PES, red shaded area, right and top axes). 
    The enhanced portions of the RIXS  spectrum relative to the PES spectrum (yellow shading) indicate regions dominated by oxygen states, including the gap edge.
    (e) O $K$-edge XAS spectrum measured in partial fluorescence yield mode at $T=20 \text{~K}$. The blue shading highlights the photon energy range from which the RIXS map is acquired in panel (a).
    }
	\label{fig:RIXS_map}
\end{figure*}

\begin{figure*}[t]
	\includegraphics[width=1\textwidth]{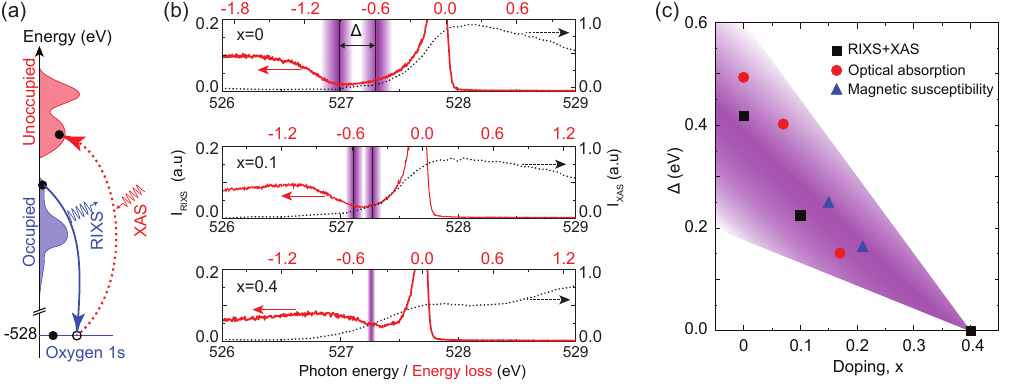} 
	\caption{Evolution of the electronic gap with potassium doping.
     (a) Cartoon sketch illustrating measurements of the unoccupied and occupied oxygen PDOS via XAS and RIXS, respectively.
     (b) Evolution of the electronic structure with doping as the indirect semiconducting gap between the occupied and the unoccupied state decreases with increasing doping. 
     (c) Dependence of the energy gap on potassium doping estimated from XAS and RIXS data, compared with gap values extracted from optical absorption and magnetic susceptibility data~\cite{kozlov,magnetic}.
     }
	\label{fig:XAS_RIXS}
\end{figure*}

\subsection*{DQMC simulations of the O displacements}
We performed \gls*{DQMC} simulations of a multi-orbital $sp$-model to study the role of \gls*{eph} coupling in establishing the persistence of local breathing distortions with  doping. The model is identical to the one used in Ref.~\cite{CohenStead2023}, which successfully describes the \gls*{CDW} insulating phase of \gls*{BBO} and its collapse with doping. Here, we consider cubic Bi$_{64}$O$_{256}$ clusters with a unit cell that includes the Bi $6s$ and O $2p_\sigma$ orbitals oriented along the three Bi-Bi bond directions. To model the O phonons and their coupling to the holes, we place a quantum harmonic oscillator on each O atom, which is polarized along the Bi-Bi bond direction and modulates the Bi-O hopping integral. We fix the corresponding phonon energy to $\hbar\Omega = 60 \text{~meV}$ and set all remaining parameters to values obtained from \gls*{DFT} calculations~\cite{Khazraie2018oxygen}. For further details, see Ref.~\cite{CohenStead2023} and the Methods section.  

Figures~\ref{fig:DQMC}(a)--(e) plot the distribution of O displacements generated during our simulations. Results are shown for $T = 300$~K and a range of hole concentrations from the parent compound $\langle n \rangle= 1$ ($x = 0$) to slightly over-doped $\langle n \rangle= 1.4$ ($x = 0.4$) in units of holes/Bi. Figs.~\ref{fig:DQMC}(f)--(j) present the same data but projected on the eigenvector of a local breathing distortion with $A_\mathrm{1g}$ symmetry 
\begin{equation}\label{eq:Xa1g}
    X_{A_\mathrm{1g}} = \frac{1}{\sqrt{6}}\left(X_{\bm{i},x}+X_{\bm{i},y}+X_{\bm{i},z}-
    X_{\bm{i},-x}-X_{\bm{i},-y}-X_{\bm{i},-z}\right).  
\end{equation}
Here, $X_{\bm{i},\pm \delta}$ denotes the displacement of the O atoms surrounding the Bi atom in unit cell $\bm{i}$. In each panel, the solid red and black lines are one- and two-Gaussian fits to the data, respectively, with the dashed lines indicating the individual components of the latter. Panels (d), (e), (i), and (j) also plot the difference between the histograms and the models in the lower half each panel. 

The distributions for $\langle n \rangle =1$ and $1.1$ each consist of a pair of equally weighted Gaussians centered symmetrically about zero displacement, indicating that the O atoms fluctuate around two inequivalent equilibrium positions. The distribution of the $A_\mathrm{1g}$ projected displacements is also bifurcated, as expected for the long-range breathing distortion predicted by this model at this temperature~\cite{CohenStead2023}. Increasing the hole concentration suppresses the long-range order, causing the two distributions to begin merging. Nevertheless, for $1.2 \le \langle n \rangle \le 1.3$, the distributions are better described as a pair of Gaussian distributions, indicating some fraction of the BiO$_6$ octahedra remains compressed in the simulations. The persistence of this bifurcation is particularly evident in the distribution of $A_\mathrm{1g}$ projected displacements, which indicates that local breathing distortions persist at these doping levels. Once the system is heavily doped ($\langle n \rangle = 1.4$), it becomes more difficult to resolve the bifurcation in $\langle X \rangle$ (Fig.~\ref{fig:DQMC}(e)), but a slight splitting can still be resolved in the projected coupling (Fig.~\ref{fig:DQMC}(j)). (This is particularly evident when one examines the differences between the fits and the data, as shown in the lower parts of Figs.~\ref{fig:DQMC}(i) and \ref{fig:DQMC}(j).) 
The observed asymmetry in the $A_\mathrm{1g}$ projected displacements is expected in a scenario where the number of bipolarons in the system is smaller than the half the number of Bi atoms in the system, as discussed in Supplementary Note 2 \cite{SuppMat}. The fact that local structural distortions are more clearly detectable in the projected measurements for $\langle n \rangle \ge 1.3$ suggests that they are fluctuating in nature. 

The trends observed in the simulated data are entirely consistent with the \gls*{PDF} results, which strongly suggests that \gls*{eph} coupling is the likely mechanism for the persistence of the local structural distortions with doping. They are also consistent with earlier \gls*{DQMC} ~\cite{Li2020quantum} and semi-classical Monte Carlo~\cite{Jiang2021polaron} simulations for related 2D models, but here for a realistic model for the 3D material. We stress that our model is derived from uncorrelated \gls*{DFT} calculations and does not include any potential enhancement of the \gls*{eph} coupling by the long-range Coulomb interaction~\cite{Yin2013correlation, Wen_ARPES_BKBO}. If we included these effects, we would likely obtain even clearer bifurcations at higher hole concentrations in the histograms. Given the relatively weak local symmetry breaking observed in Fig.~\ref{fig:refinement}(c), it may be possible to place an upper bound on the strength of the \gls*{eph} coupling, which would provide valuable insights into the nature of pairing at lower temperatures.

\subsection*{Signatures of bond disproportionation} 
\gls*{RIXS} experiments were performed on thin films of \gls*{BKBO} grown by \gls*{PLD}. By using thin film samples, we were able to overcome the challenge of synthesizing sufficiently large single crystals of \gls*{BKBO} in the under- to optimally-doped regime ($x = 0$ to $0.4$) and avoid the spectral degradation that might otherwise occur due to the surface roughness of cleaved crystals. The large surfaces of the films also facilitate beam rastering to mitigate potential sample damage. 

Figure~\ref{fig:RIXS_map}(a) maps the parent ($x=0$) compound's \gls*{RIXS} spectra at $T = 20$ K as a function of the incident photon energy. The excitation energy is tuned over the lowest-lying O $K$-edge peak measured by \gls*{XAS}, as shown in Fig.~\ref{fig:RIXS_map}(e). The \gls*{RIXS} spectra consist of two main components exhibiting distinct behaviors as a function of the excitation energy. 
Supplementary Note 3 \cite{SuppMat} provides a sketch of the relevant \gls*{RIXS} processes \cite{Ament2011resonant, Ament2011, Bieniasz2021, Gilmore2023quantifying}. 
The first and most obvious component is the prominent, broad peak centered roughly around -4 eV energy loss, with a weaker flat feature extending to an energy loss of about -1 eV. Increasing the incident photon energy directly shifts this portion of the \gls*{RIXS} spectrum to higher energy loss (as seen in Fig.~\ref{fig:RIXS_map}(a)), indicating that these are valence band fluorescence features. The second key component of the spectrum, highlighted in Figs.~\ref{fig:RIXS_map}(b)--(c), is an asymmetric broadening near the elastic line, which is a signature of low-energy excitations, such as excitations of the lattice due to \gls*{eph} coupling~\cite{Ament2011, Lee2013, Devereaux2015, Bieniasz2021}. This feature is most evident at an excitation energy of 527.9 eV (red curve in Fig.~\ref{fig:RIXS_map}(c)), where the inelastic portion of the spectrum appears largest compared to the elastic peak. 
 
The fluorescence features provide a useful window into \gls*{BKBO}'s electronic structure that affirms the relevance of negative charge transfer energies in the system by showing that the gap opens in predominantly oxygen-derived states~\cite{Foyevtsova2015}. Fig.~\ref{fig:RIXS_map}(d) compares the \gls*{RIXS} spectrum of \gls*{BBO} with data from angle-integrated \gls*{PES}. The spectra are aligned and scaled to roughly match features in the high energy loss/deep binding energy region. The \gls*{PES} data was acquired from an \textit{in situ} \gls*{BBO} film, as described in Ref.~\cite{Plumb2016}. The \gls*{PES} signal derives from the \gls*{DOS}, albeit modulated by photoexcitation matrix elements \cite{Damascelli2004}. The \gls*{RIXS} fluorescence spectrum, on the other hand, is due to the decay of O $2p$ states to fill the O $1s$ core hole and thus relates to the oxygen \gls*{PDOS}. The overall similarities between these two spectra highlight the profound degree of oxygen hybridization throughout the band structure~\cite{Foyevtsova2015, Plumb2016, Khazraie2018oxygen}. 
The differences in the spectra of Fig.~\ref{fig:RIXS_map}(d) are just as revealing. In particular, compared to the photoemission spectrum, the \gls*{RIXS} fluorescence exhibits an enhanced shelf-like protrusion in the region of about -1 to -3 eV energy loss. The band dispersion of these highest occupied states is seen in an angle-resolved view of the photoemission spectrum \cite{Plumb2016}, but when the data is angle- (i.e., momentum-) integrated to obtain the \gls*{PES} spectrum, the intensity of this feature is almost totally drowned out by the height of the main peak near -3.5 eV binding energy. The pronounced appearance of the shelf-like feature in \gls*{RIXS} compared to \gls*{PES} implies a resonant enhancement, indicating that oxygen states dominate at the gap edge. A similar resonant enhancement can be seen on the right side of the main peak. Both enhancements are consistent with the calculated oxygen \gls*{PDOS} of \gls*{BBO} \cite{Foyevtsova2015}. Hence, the \gls*{RIXS} fluorescence paints a picture of strong Bi-O hybridization accompanied by a gap opening in predominantly O-derived states, consistent with the model of a bond-disproportionated phase generated by a reverse transfer of holes onto the ligands \cite{Foyevtsova2015, Khazraie2018oxygen}.

\begin{figure}[t]
    \centering
    \includegraphics[width=\columnwidth]{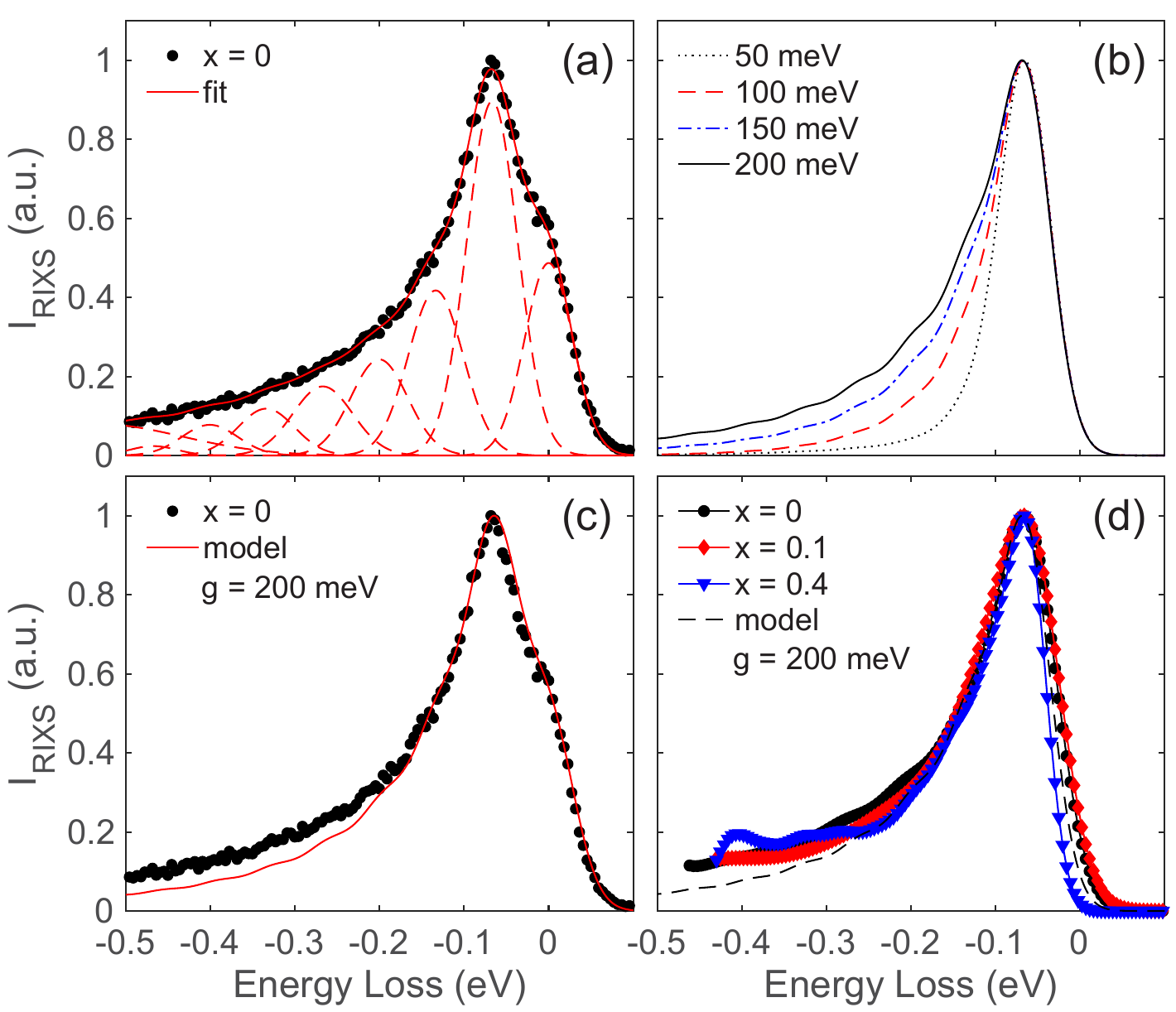}
    \caption{Probing the $e$-ph coupling with oxygen $K$-edge RIXS.
    (a) Fitted phonon harmonics for the BBO ($x = 0$) sample. The dashed lines show the contributions from each phonon harmonic while the solid line is the sum. 
    (b) Calculated RIXS intensity obtained from our cluster calculations as a function of the $e$-ph coupling strength $g$. Increasing $g$ results in a longer tail of lattice excitations extending to higher energy loss values. 
    (c) Comparison of O $K$-edge RIXS spectra of the parent compound BBO and the predictions of our model for $g = 200$ meV. 
    (d) Comparison of the measured O $K$-edge RIXS spectra in BKBO for different values of $x$. All spectra have a similar shape with a significant tail extending to higher energy loss, indicative of a strong $e$-ph coupling. The black dashed line in this panel corresponds to the same model data plotted in panel (c). The elastic line has been removed from the 
    experimental data points to focus on the phonon features.
    }
    \label{fig:RIXS_theory}
\end{figure}

These insights are reinforced by a comparison of the \gls*{XAS} spectrum with the fluorescence emission in \gls*{RIXS}. As illustrated schematically in Fig.~\ref{fig:XAS_RIXS}(a), the 
total fluorescence yield \gls*{XAS} intensity is due to the transition from the O $1s$ core level to the O $2p$ states, thus relating to the oxygen PDOS of the \textit{unoccupied} bands. On the other hand, as noted earlier, the fluorescence intensity of the \gls*{RIXS} spectrum stems from electron decay from the $2p$ bands to empty core level states, consequently tracking the oxygen PDOS of the \textit{occupied} bands. Thus, the \gls*{XAS} and \gls*{RIXS} fluorescence spectra complement each other and provide a means to follow the evolution of the indirect bandgap as a function of doping.

Figure~\ref{fig:XAS_RIXS}(b) displays the indirect gap $\Delta$ estimated from the \gls*{XAS} and \gls*{RIXS} spectra for $x = 0$, $0.1$, and $0.4$ (see also Supplementary Note 3 \cite{SuppMat}). 
Upon hole doping, the dominant absorption peak shifts to lower photon energy, while the leading edge of the fluorescence in the \gls*{RIXS} spectra shifts to lower  (i.e., closer to zero) energy loss. 
The estimated gap between the \gls*{XAS} and \gls*{RIXS} fluorescence spectra decreases with doping, as summarized in Fig.~\ref{fig:XAS_RIXS}(c), and accounts for the change in the indirect gap of the oxygen bands. The changes in the band's filling and the gap between the \gls*{XAS} and \gls*{RIXS} fluorescence leading edge are reasonably close to the indirect gap estimates obtained from optical absorption and magnetic susceptibility measurements~\cite{UWE, kozlov, Hellman1990, magnetic}. The fact that the gap closure is observed via the oxygen \gls*{PDOS} further affirms the dominant role of O $2p$ states at the gap edge, which closes due to the hole doping introduced by the potassium substitution. 

\subsection*{Probing the \texorpdfstring{$e$}{e}-ph coupling with RIXS}
Figure~\ref{fig:RIXS_theory} analyzes the near-elastic component of the \gls*{RIXS} spectra to study the \gls*{eph} coupling as a function of doping. The data were acquired at $T = 20$ K with incident photon energy tuned to the maximum of the lowest-lying O $K$-edge absorption, as was shown in Fig.~\ref{fig:RIXS_map}. 
We can decompose the spectra for each doping level into an elastic component and a single phonon excitation at $\hbar\Omega \approx 65~\text{meV}$ and its overtones We illustrate this by fitting the parent compound spectrum with a series of weighted harmonics in Fig.~\ref{fig:RIXS_theory}(a). (Supplementary Note 3 shows the decomposition for the other doping levels \cite{SuppMat}. There we also note that we observe a very weak temperature dependence in of the spectra in the quasi-elastic region.)

To gain more insight into the nature of these harmonic phonon excitations, we use exact diagonalization to calculate the low-energy \gls*{RIXS} spectrum of a BiO$_6$ octahedron with open boundary conditions within the Kramers-Heisenberg formalism. To reduce the Hilbert space associated with orbital and phonon subspaces, our model includes the Bi $6s$ orbital, the $A_\text{1g}$ combination of ligand oxygen orbitals, and the corresponding local breathing mode, which directly modulates the Bi-O hybridization (see Methods). All model parameters are identical to those used in our \gls*{DQMC} calculations, except the phonon energy $\hbar\Omega$ and \gls*{eph} coupling $g$, which we now treat as fitting parameters. We further fix the inverse core hole lifetime parameter $\Gamma = 150~\text{meV}$, which is typical for the O $K$-edge~\cite{Kotani1979, Lee2013}. 

As shown in Fig.~\ref{fig:RIXS_theory}(b), the magnitude of the \gls*{eph} coupling in our model determines the relative intensities of the phonon harmonics. We can thus estimate the coupling strength's value and doping dependence from the overall shape of the spectrum's tail, with some important caveats. Because we are approximating the system with a small local cluster, our approach is very similar in spirit to the single-site framework introduced by Ament {\it et al}.~\cite{Ament2011} but extended to off-diagonal \gls*{eph} interactions. Small cluster approaches like these neglect charge and orbital fluctuations beyond the cluster, which can introduce additional momentum dependence to the spectra~\cite{Bieniasz2021}. Our approach also does not account for coupling to multiple phonon modes~\cite{Geondzhian2020generalization} or any potential coupling between the lattice and the core-hole~\cite{Geondzhian2018demonstration}, which might affect the shape of the spectra and estimates for the \gls*{eph} coupling. 

We obtain good agreement with the data for a microscopic coupling of $g = 200~\text{meV}$, as shown in Fig.~\ref{fig:RIXS_theory}(c). 
In particular, the overall line shape of the phonon excitations is well reproduced by the model which only couples to the local breathing model, with slight deviations appearing in energy losses above the third harmonic. (We attribute this discrepancy to charge fluctuations beyond the length scale of the cluster.) 
The microscopic coupling extracted from the fits is a factor of two larger than the one inferred from first principles calculations, which we attribute to limitations in the model. Nevertheless, the results suggest that the oxygen holes in \gls*{BBO} are coupled strongly to the breathing motion of the oxygen atoms. Crucially, we observe nearly identical profiles for the lattice excitations in the doped samples. Fig.~\ref{fig:RIXS_theory}(d) illustrates this by overlaying the \gls*{RIXS} spectra for $x = 0$, $0.1$, and $0.4$ samples together with the model data. Here, we have subtracted the contribution from the elastic line and normalized each spectrum to the intensity at the first phonon line to enable direct comparisons. Since the relative intensities of the phonon harmonics (and hence the overall shape of each spectrum) are determined by the strength of the \gls*{eph} coupling, this result suggests that substantial coupling to the breathing distortions persists with hole doping.

\section*{Discussion} 
We have used resonant inelastic x-ray and neutron total scattering techniques to study the evolution of \gls*{BKBO}'s local crystal structure, electronic structure, and \gls*{eph} coupling with doping. Analysis of the neutron \gls*{PDF} finds evidence that local BiO$_6$ breathing distortions --- residuals of the parent insulating phase --- persist deep into the metallic phase, even at the highest measured K concentration ($x=0.4$), which is at or slightly beyond the optimal doping level for superconductivity. Details of the O $K$-edge \gls*{RIXS} measurements affirm the influence of an effective negative charge transfer energy in \gls*{BKBO}, which drives the doped holes to reside primarily on the O $2p$ orbitals.  Low-energy excitation features in the \gls*{RIXS} data indicate that significant coupling between the oxygen holes and the Bi-O bond-stretching phonons exists throughout the investigated doping range. Small cluster \gls*{ED} and state-of-the-art \gls*{DQMC} simulations on extended lattices demonstrate that our measurements can be explained using a multi-orbital model in which the oxygen vibrations directly modulate the Bi-O hybridization. Our results thus provide compelling evidence for the presence of small BiO$_6$ (bi)polarons in \gls*{BKBO}'s metallic phase, which persist through optimal doping. 

The findings from our neutron scattering experiments stand in contrast to a recent x-ray \gls*{PDF} study of \gls*{BKBO}~\cite{Greven2023}, which reported that short-range breathing distortions already vanish at finite doping levels below the insulator-metal transition, and instead found evidence of local displacements of the Bi atoms along the Bi-O axis that break inversion symmetry. Such findings, if verified, would severely constrain the relevance of (bi)polaron interactions for superconductivity in the bismuthates. 
We have been unable to reconcile our data with the conclusions of Ref.~\cite{Greven2023}. Not only do we observe the persistence of breathing distortions at doping levels where superconductivity occurs, but we have also not been able to detect the loss of an inversion center in our \gls*{PDF} data. When attempting to break inversion symmetry, the refinements fail to converge due to significant, and in some cases complete, correlations between the atomic position parameters that correspond to single parameters in the centrosymmetric model. 

At present, we cannot explain definitively why our findings diverge so dramatically from Ref.~\cite{Greven2023}, but the experiments differ in some notable respects. Compared to neutrons, the sensitivity of hard x-rays to the crucial O atoms is vastly diminished, and the O ($Z=8$) scattering contribution is dwarfed by the signals from Ba ($Z=56$) and Bi ($Z=83$). 
Another difference is that hard x-ray scattering should have a sizeable inelastic component, while the thermal neutrons in our measurements scatter totally elastically. The use of hard x-rays also carries a risk of radiation-induced sample damage. Furthermore, while our neutron experiments were performed on powder samples obtained by solid state synthesis, the x-ray \gls*{PDF} measurements were performed on single crystals grown by an electrochemical method, which in some studies have exhibited significant doping inhomogeneity~\cite{Minami1996}.  Finally, the coherent neutron scattering lengths for Ba and K are short, which reduces our sensitivity to these atoms. If an inversion symmetry breaking is driven by ordering of these cations, rather than within the BiO$_6$ octahedra, it is possible our experiments would not detect it. 
Nevertheless, this would not impact our conclusion that local breathing distortions are present up to high doping levels $x \le 0.4$.

The bismuthates are classified as negative charge transfer systems with significant hybridization between the Bi $6s$ orbitals and ligand oxygen orbitals~\cite{Foyevtsova2015, Khazraie2018oxygen, Khazraie2018bond}. However, other families of perovskites share similar characteristics, whereby a lattice-coupled transfer of hole pairs onto the ligands should occur \cite{Park2012site, Johnston2014charge, Bisogni2016groundstate, Dalpian2018, Benam2021}. Our results are likely relevant to this much broader class of materials. For example, the notion that bipolarons may condense into insulating or superconducting states provides a natural framework for understanding these systems' unusual metal-insulator transitions, non-Fermi-liquid transport, 
and superconductivity. Here, the nature of the broken symmetry state at low temperatures would depend on the relative strengths of the cation-ligand hopping, charge transfer energy, and density of holes. For instance, in rare earth nickelates, which undergo metal-insulator transitions, recent neutron total scattering experiments found similar evidence of local breathing distortions~\cite{Li2016insulating, Shamblin2018}, and dielectric spectroscopy \cite{Shamblin2018} and transport \cite{Tyunina2023small} measurements find evidence for a polaronic liquid metallic phase. In another likely related system, the recently discovered superconductivity in Ba$_{1-x}$K$_x$SbO$_3$ arises out of an ordered insulator phase in a similar manner as in \gls*{BKBO} \cite{Kim2022superconductivity}. However, the insulating phase in the antimonate persists to larger hole concentrations, probably due to differences in the charge transfer energy and Sb-O hybridization. 
Our results thus call for systematic studies of these materials focusing on local structural properties and \gls*{eph} coupling.

\section*{\label{sec:methods}Methods}

\subsection*{Neutron scattering experiments}
The neutron scattering measurements were performed on polycrystalline \gls*{BKBO} samples synthesized via solid-state reaction methods. 
\Gls*{XRD} measurements at room temperature confirmed the samples' phase. 
The atomic ratios in the various \gls*{BKBO} samples were measured by \gls*{XRF}. Both \gls*{XRD} and \gls*{XRF} indicate successful and homogeneous Ba/K substitution in all the studied samples. Additional sample characterization is provided in Supplementary Note 4 \cite{SuppMat}.

Neutron total scattering experiments were performed on the \gls*{NOMAD} at the \gls*{SNS} at \gls*{ORNL}~\cite{Neuefeind2012}. All samples were placed in quartz capillaries (2 mm diameter, 0.01 mm wall thickness) and exposed to the neutron beam for 48 minutes in two separate scans of 24 minutes each. PDF analysis of each individual scan, as well as their final sum for each sample, was used to verify the statistical robustness of the results in the low-$r$ region of interest (see Supplementary Note 1 \cite{SuppMat}). Calibration of \gls*{NOMAD}'s time-of-flight detectors was performed by scattering from diamond powder. The structure function $S(Q)$ was obtained by subtracting the background (an empty quartz capillary) from the sample measurement and normalizing to scattering from a vanadium rod to account for the neutron spectrum and detector effects. The experimental pair distribution $G(r)$ was extracted via the Fourier transform
\begin{equation}\label{eq:fourier}
    G(r)=\frac{2}{\pi}\int Q[S(Q)-1]\sin(Qr)~dQ ,
\end{equation}
where the scattering vector $Q$ ranged from 0.2 to 31.4~$\textrm{\r{A}}^{-1}$. The GSAS-II program \cite{Toby2013} was used to perform Rietveld refinement on bank 3 of the neutron diffraction patterns. An instrument parameter file was created using neutron diffraction data collected from Si powder. The PDFGui program \cite{Farrow2007} was used for small-box refinement of the pair distribution functions. Refined variables for structural analysis include the global scale factor, unit cell parameters, and all symmetry-permitted atomic translational movement and displacement parameters. The goodness-of-fit is calculated via the relation
\begin{equation}
    R^2_w= \frac{ \sum_{i=1}^{N} w(r_i)[ G_\mathrm{obs}(r_i)-G_\mathrm{calc}(r_i)]^2 }{\sum_{i=1}^{N} w(r_i)[G_\mathrm{obs}(r_i)]^2},
    \label{FT_PDF}
\end{equation}
where $w$ represents the weight of the data points, $G_\mathrm{obs}$ is the experimental data point, and $G_\mathrm{calc}$ is the simulated data point. All small-box refinements were performed over the range $r = 1.5$ to 10~\AA.

\subsection*{RIXS experiments}
\Gls*{RIXS} experiments were performed on \gls*{BKBO} thin films grown by \gls*{PLD}, as these samples provide sufficient size, quality, and surface flatness over the investigated doping range. 
\Gls*{BKBO} films with a thickness of 12 nm were deposited on SrTiO$_3$(001) substrates from ablation targets with doping levels of $x=0$, $0.1$, and $0.4$. The growth conditions are described in previous work \cite{Plumb2016, naamneh:2018}. The growth was epitaxial along the $c$-axis, as demonstrated by \gls*{RHEED} measurements. The pseudocubic lattice parameter at each doping level was measured by room temperature \gls*{XRD}. 
Resistivity measurements show that $x=0.4$ film is superconducting below $T_\mathrm{c}=20 \text{~K}$. The lower-doped samples are insulating. Despite the fact that the $x=0.4$ film's $T_\mathrm{c}$ is lower than in powder samples, structural, spectroscopic, and \textit{normal state} transport measurements indicate that the doping levels of the thin films and powders are well-matched, and that the two sets of samples align at very similar positions along the $x$ axis of the electronic phase diagram. Details of the thin film characterizations are provided in the Supplementary Note 4 \cite{SuppMat}.

The \gls*{RIXS} experiments were performed at the ADRESS beamline of the Swiss Light Source \cite{Strocov}. The samples were transferred from the film deposition chamber to the \gls*{RIXS} system via a vacuum suitcase with pressure better than $10^{-7}$ mbar to minimize surface contamination and preserve the surface quality. The sample temperature during the experiments was 17 K unless otherwise noted. The energy resolution was determined to be 70 meV by measuring the \gls*{FWHM} of the elastic line on carbon tape. All \gls*{RIXS} measurements were performed at the oxygen $K$-edge under fixed geometry corresponding to the scattering wavevector $\bm{q}=(0.24,0,0.17)$ r.l.u. The positioning of the sample was shifted after each measurement to minimize the risk of damage from irradiation.

\subsection*{RIXS calculations}
The lattice excitations in the \gls*{RIXS} spectra were modeled within the Kramers-Heisenberg formalism, where the initial $\ket{i}$, intermediate $\ket{m}$, and final states $\ket{f}$ of the scattering process are obtained from \gls*{ED} calculations. 
We approximate the system in the insulating phase of \gls*{BBO} with a single BiO$_6$ octahedron. The cluster Hamiltonian, written in hole-language, is $H = H_e + H_\mathrm{ph} + H_{e-\text{ph}}$, 
where 
\begin{equation}
    \begin{aligned}
        H_e&=\epsilon_s \sum_\sigma s^\dagger_{\sigma}s^{\phantom\dagger}_{\sigma} + 
        \epsilon_p \sum_{\sigma,\nu} p^\dagger_{\nu,\sigma}p^{\phantom\dagger}_{\nu,\sigma}\\
        &+t_{sp} \sum_{\sigma,\nu} \left(Q_\nu s^\dagger p_{\nu,\sigma} + h.c.\right)
    \end{aligned}
\end{equation}
describes the electronic sector, 
\begin{equation}
    H_\text{ph} = \hbar\Omega \sum_\nu \left(b^\dagger_\nu b^{\phantom\dagger}_\nu + \tfrac{1}{2}\right)
\end{equation}
describes the phononic sector and 
\begin{equation}
    H_{e-\text{ph}} = g\sum_{\nu,\sigma}\left[s^\dagger p_{\nu,\sigma}(b^\dagger_\nu + b_\nu) + \text{h.c.}\right], 
\end{equation}
describes the \gls*{eph} coupling due to the modulation of the Bi-O hopping integrals. Here, $s^\dagger_{\sigma}$ creates a spin-$\sigma$ hole on the Bi $6s$ orbital; $p^\dagger_{\nu,\sigma}$ creates a hole in each the O $2p_\nu$ orbital ($\nu = \pm x, \pm y, \pm z$) oriented toward the Bi atom; $b^\dagger_\nu$ creates a phonon on the ligand oxygen site $\nu$; $\epsilon_s$ and $\epsilon_p$ are the on-site energies of the $6s$ and $2p$ orbitals, respectively; $\hbar\Omega$ is the phonon energy; $t_{sp}$ is the Bi-O hopping integral with phase factors $Q_{\pm y} = Q_{\pm y} = Q_{\pm z} = \mp 1$ (see Fig. 1a of Ref. \cite{CohenStead2023}). For simplicity, we have neglected the O-O hopping as it does not play a major role in the resulting model. 

Next, we introduce a molecular orbital basis for the O orbitals~\cite{Khazraie2018bond}
\begin{equation}
    \begin{aligned}
        L_s &= \frac{1}{\sqrt{6}}\left(p_{x}+p_{y}+p_{z}-p_{-x}-p_{-y}-p_{-z}\right), \\
        L_x &= \frac{1}{\sqrt{2}}\left(p_{x}+p_{-x}\right), \\
        L_y &= \frac{1}{\sqrt{2}}\left(p_{y}+p_{-y}\right), \\
        L_z &= \frac{1}{\sqrt{2}}\left(p_{z}+p_{-z}\right), \\
        L_{x^2-y^2} &= \frac{1}{2}\left(p_{x}-p_{y} - p_{-x}+p_{-y}\right),~\text{and{}} \\
        L_{3z^2-r^2} &= \frac{1}{2\sqrt{3}}\left(p_{x}+p_{y}-2p_{z}-p_{-x}-p_{-y}+2p_{-z}\right), 
    \end{aligned}
\end{equation}
where we have suppressed the spin index for brevity. We can define analogous transformations for the phonon operators. For example, the annihilation operator for the breathing mode is given by 
\begin{equation*}
    B_s = \frac{1}{\sqrt{6}}\left(b_{x}+b_{y}+b_{z}-b_{-x}-b_{-y}-b_{-z}\right). 
\end{equation*}
This operator corresponds to the local breathing distortion with $A_\mathrm{1g}$ symmetry, as defined in Eq.~\eqref{eq:Xa1g}.  When written in this molecular basis, the cluster Hamiltonian simplifies to 
\begin{equation}
    \begin{aligned}
        H&= \epsilon_s \sum_\sigma s^\dagger_{\sigma}s^{\phantom\dagger}_{\sigma} + 
        \epsilon_p \sum_{\sigma,\nu} L^\dagger_{\nu,\sigma}L^{\phantom\dagger}_{\nu,\sigma} + \Omega\sum_\nu \left(B^\dagger_{\nu} B^{\phantom\dagger}_\nu + 1/2\right) \\
        &-\sqrt{6}t_{sp} \sum_{\sigma} \left(s^\dagger L_{s,\sigma} + h.c.\right) \\
        &+g\sum_{\nu,\sigma} \left[s^\dagger_\sigma L^{\phantom\dagger}_{\nu,\sigma} \left(B^\dagger_\nu + B^{\phantom\dagger}_\nu\right)+ h.c.\right], 
    \end{aligned}\label{eq:Hmolecular}
\end{equation}
where the index $\nu = s, x, y, z, x^2-y^2$, and $3z^2-r^2$. 
We now further reduce the system to an effective two-level problem by neglecting all molecular orbitals that don't hybridize directly with the Bi orbitals  
\begin{equation}
    \begin{aligned}
        H&= \epsilon_s \sum_\sigma s^\dagger_{\sigma}s^{\phantom\dagger}_{\sigma} + 
        \epsilon_p \sum_{\sigma} L^\dagger_{s,\sigma}L^{\phantom\dagger}_{s,\sigma} + \hbar\Omega\left(B^\dagger_{s} B^{\phantom\dagger}_s + \tfrac{1}{2}\right)\\ 
        &+\sqrt{6}t_{sp} \sum_{\sigma} \left(s^\dagger L_{s,\sigma} + h.c.\right)\\
        &+g\sum_{\sigma} \left[s^\dagger_\sigma L_{s,\sigma} \left(B^\dagger_s + B_s\right)+ h.c.\right].
    \end{aligned}\label{eq:Heff}
\end{equation}
Eq.~\eqref{eq:Heff} can now be easily diagonalized while retaining a large number of phonon quanta $N_\mathrm{max}$ for the oxygen breathing mode. 

We compute the \gls*{RIXS} intensity at the O $K$-edge using the Kramers-Heisenberg formula 
in this orbital basis
\begin{equation}\label{eq:Irixs}
    I(\omega) = \sum_{f} \left\vert M_{f,g}\right\vert^2 \delta(E_f - E_i - \omega)
\end{equation}
where $\omega$ is the energy transfer, 
\begin{equation}\label{eq:ME}
    M_{fg} = \sum_\sigma \frac{\bra{f}L^\dagger_\sigma \ket{n}\bra{n}L^{\phantom\dagger}_\sigma \ket{g}}{E_g-E_n+\omega_\mathrm{in}+\mathrm{i}\Gamma/2}  
\end{equation}
is the scattering amplitude, $\ket{g}$, $\ket{n}$, and $\ket{f}$ are the initial, intermediate, and final states of the scattering process, and $\Gamma/2$ is the inverse core-hole lifetime. 

For numerical calculations, we set (in units of eV) $t_{sp} = 2.31$, $\epsilon_s = 6.23$, $\epsilon_p = 4.14$ based on \gls*{DFT} calculations \cite{Khazraie2018oxygen, CohenStead2023}. When evaluating Eq.~\eqref{eq:ME}, we tuned the incident photon energy to the location of the maximum in the computed \gls*{XAS} and adopted $\Gamma/2 = 150$ meV~\cite{Lee2013}. The initial and final states are obtained by diagonalizing Eq.~\eqref{eq:Heff} in the two-hole, $m_z = 0$ sector while retaining $N_\mathrm{max} = 100$ phonon quanta. The intermediate states are obtained by diagonalizing Eq.~\eqref{eq:Heff} in the one-hole sector and neglecting the influence of the O $1s$ core hole, which only produces a trivial shift in the eigenvalue spectrum at our level of modeling. The energy-conserving $\delta$-function in Eq.~\eqref{eq:Irixs} was approximated with a Gaussian line shape with a standard deviation of $\sigma = 30$ meV, as set by the leading edge of the elastic line. Since the overall weight of the elastic line is determined by extrinsic factors like surface roughness, we re-scaled its contribution in our calculation by a factor of 0.17 to reproduce the experimental weight and normalized the spectra to their intensity at the first phonon line to compare to the data. The \gls*{RIXS} data is best reproduced for $\hbar\Omega = 65$ meV and $g = 0.2$ eV, with the latter being a factor of two larger than the value inferred from \gls*{DFT} calculations~\cite{Meregalli1998electron, CohenStead2023}. This difference is not surprising; a small cluster suppresses band effects, which can impact the intensity of lattice excitations~\cite{Bieniasz2021}. 

\subsection*{DQMC Simulations} 
The \gls*{DQMC} simulations of the \gls*{BKBO} models were performed on $L=4$ lattice at $T=300~\text{K}$ using the \texttt{SmoQyDQMC.jl} package~\cite{SmoQy1, SmoQy2}. Simulations were run using 12 parallel walkers, each performing 5,000 \gls*{HMC} updates to thermalize the system, followed by an additional 10,000 updates, after each of which measurements were made. The \gls*{HMC} update consisted of $N_t = 3$ time-steps with corresponding step-size $\Delta t = \pi/(2 N_t \Omega)$. The chemical potential was dynamically updated during the simulation to achieve the desired hole density using the algorithm presented in Ref.~\cite{Miles2022dynamical}.

\section*{Data Availability}
The data supporting this study are available at Zenodo:
 \url{ https://doi.org/10.5281/zenodo.17280978}

\section*{Code Availability} 
A Julia notebook for the \gls*{RIXS} calculations is available at Zenodo:
\url{ https://doi.org/10.5281/zenodo.17280978}

The \texttt{SmoQyDQMC.jl} package can be obtained at \url{https://github.com/SmoQySuite/SmoQyDQMC.jl}.

\section*{Acknowledgements}
The authors thank M.~Berciu, J.~Chang, J.~Neuefeind, G.~A.~Sawatzky, and Q.~Wang for helpful discussions. M.~N.~was supported by the ISRAEL SCIENCE FOUNDATION (grant No.~2509/20). E.~O'Q., M.~L., and S.~J. were supported by the National Science Foundation Materials Research Science and Engineering Center program through the UT Knoxville Center for Advanced Materials and Manufacturing (DMR-2309083). M.~N., W.~R.~P., and N.~C.~P. were supported by the Swiss National Science Foundation through Project Nos.~200021\_159678 and 200021\_185037. E.~P., D.~M., Y.~T., and T.~S. were supported by the Swiss National Science Foundation through the NCCR MARVEL (grant number 141828) and the Sinergia project Mott Physics Beyond the Heisenberg Model (MPBH) (grant number 160765). The synchrotron experiments were performed at the ADRESS beamline of the Swiss Light Source at the Paul Scherrer Institut (PSI). The neutron scattering experiments used resources at the Spallation Neutron Source, a DOE Office of Science User Facility operated by the Oak Ridge National Laboratory.

\section*{Author Contributions} 
J.~S., E.~C.~O'Q., and M.~K.~L.~performed neutron scattering experiments. M.~N., E.~P., D.~M., Y.~T., W.~P., M.~R., N.~C.~P., and T.~S.~performed \gls*{RIXS} experiments. D.~J.~G.~synthesized and characterized powder samples for the neutron experiments, as well as ablation target material for the thin films studied by \gls*{RIXS}. M.~N.~and M.~R.~grew and characterized the thin film samples. B.~C-.S.~performed \gls*{DQMC} simulations. S.~J.~performed \gls*{RIXS} calculations. M.~N., E.~C.~O'Q., S.~J., and N.~C.~P.~wrote the manuscript with input from all authors.

\bibliography{references.bib}

\end{document}